\documentclass[journal]{IEEEtran}
\usepackage{cite}
\usepackage{amsmath,amssymb,amsfonts}
\usepackage{algorithm}
\usepackage{algorithmic}
\usepackage{graphicx}
\usepackage{textcomp}
\usepackage{booktabs}
\usepackage{array}
\usepackage{bm}
\usepackage{hyperref}

\def\BibTeX{{\rm B\kern-.05em{\sc i\kern-.025em b}\kern-.08em
    T\kern-.1667em\lower.7ex\hbox{E}\kern-.125emX}}
    
\begin{document}
\title{Flow Matching-Based\\PET Image Reconstruction}
\author{Fumio Hashimoto, Ziqian Huang, Tatsuya Yokota, and Kuang Gong
\thanks{This work was supported by NIH grants R01EB034692 and R01AG078250. (Corresponding author: Kuang Gong)}
\thanks{F. Hashimoto, Z. Huang and K. Gong are with the J. Crayton Pruitt Family Department of Biomedical Engineering, University of Florida, FL, USA (e-mail: fumio.hashimo@ufl.edu; ziqian.huang@ufl.edu; kgong@bme.ufl.edu).}
\thanks{T. Yokota is with the Department of Computer Science, Nagoya Institute of Technology, Aichi, Japan and with the RIKEN Center for Advanced Intelligence Project, Tokyo, Japan (e-mail: t.yokota@nitech.ac.jp).}}

\maketitle

\begin{abstract}
Generative models have shown strong potential for positron emission tomography (PET) image reconstruction. Although diffusion model-based reconstruction methods have demonstrated promising performance, they often require many reverse sampling steps with data-consistency updates incorporated into the sampling process. Flow matching offers an attractive alternative because it can directly estimate clean images from intermediate states, allowing data-consistency refinement to be separated from flow propagation. In this work, we proposed flow matching-based PET image reconstruction methods. We first established PET-FlowDPS by incorporating Poisson likelihood guidance with an expectation-maximization (EM)-based preconditioner into the FlowDPS framework. We then proposed a model-based PET reconstruction method that used a pretrained flow matching model as a prior, in which the flow-based prior, PET data refinement, and stochastic propagation were interpreted within an approximate Bayesian framework. Experimental results using [$^{\text{18}}\text{F}$]FDG brain PET datasets showed that the proposed method achieved better bias-variance trade-offs across different dose levels compared with other reference methods. These results demonstrated the potential of flow matching as a generative prior for quantitative PET image reconstruction.
\end{abstract}


\section{Introduction}
\label{sec:introduction}
\IEEEPARstart{P}{ositron} emission tomography (PET) enables quantitative \textit{in vivo} imaging of radiotracer distributions, making it a valuable molecular imaging modality in oncology, cardiology, and neurology. However, PET image quality is fundamentally constrained by limited counting statistics and various physical degradation effects, which compromises its lesion detectability and quantitative accuracy. Although iterative reconstruction methods, such as maximum-likelihood expectation-maximization (MLEM) and ordered-subsets expectation-maximization (OSEM), have been widely used for PET image reconstruction with explicit modeling of Poisson noise statistics and the PET system response~\cite{Shepp1982,Hudson1994}, the resulting images often exhibit excessive noise under low-count conditions. To address this limitation, maximum \textit{a posteriori} (MAP)-based reconstruction methods have been developed to incorporate prior information, such as spatial smoothness and anatomical information, to further improve PET image quality~\cite{Green1990,DePierro1995,Nuyts2003,Qi2006,Qi2000}.

Recent advances in deep learning have led to growing interest in data-driven approaches for PET image reconstruction~\cite{Reader2020,Hashimoto2024Review,miwa2025innovations}. Early deep learning-based methods mainly used convolutional neural networks (CNNs) as learned image priors within PET image reconstruction. For example, CNNs have been used to parameterize PET images within iterative reconstruction frameworks~\cite{Gong2019} and to provide learned regularization in unrolled reconstruction frameworks~\cite{Mehranian2021}. Although these approaches demonstrate the potential of CNN-based priors to improve PET image quality, they rely on deterministic mappings rather than explicitly modeling the underlying image distribution. When multiple plausible high-quality solutions exist, such deterministic mappings may produce averaged estimates, potentially suppressing fine structural details and leading to oversmoothed images~\cite{Whang_2022_CVPR,Saharia2023}. 

Diffusion models~\cite{Ho2020,Song2021SDE}, which can learn image distributions and act as generative priors, have been investigated for various PET imaging tasks, including image denoising~\cite{Gong2024DDPM,Pan2024PETCM,Yu2025DDPM,Xie2026DDPET}, synthesis~\cite{Xie2024,Yuxin2026}, and attenuation correction~\cite{Chen2025AC,jeong2026gpdm}. For PET image reconstruction, Singh et al.~\cite{singh_2024} demonstrated the potential of score-based reconstruction using diffusion posterior sampling (DPS)~\cite{chung2023diffusion} and decomposed diffusion sampling (DDS)~\cite{chung2024decomposed}. Subsequent studies extended diffusion-based PET reconstruction to fully 3D imaging\cite{webber2025lisch}, joint activity and attenuation estimation~\cite{bae2025joint,PhungMLAA}, and out-of-distribution adaptation~\cite{Hashimoto2026,yilmaz2026petadaptertesttimedomainadaptation}.

Despite these advances, diffusion-based PET image reconstruction remains computationally demanding because both neural network evaluations and data-consistency updates must be repeated over many reverse sampling steps. Its performance can also be sensitive to the noise schedule, the number of reverse sampling steps, and the strength and timing of the data-consistency updates. In DPS, computing the likelihood gradient with respect to an intermediate noisy image requires differentiating through the denoised estimate and thus backpropagating through the neural network at each sampling step. This increases computational and memory costs and may result in numerically sensitive updates~\cite{chung2024decomposed}. Although PET-DDS~\cite{singh_2024} and likelihood scheduling~\cite{webber2025lisch} improve the efficiency and robustness of data-consistency updates, the generative prior and data-consistency refinement remain coupled within the prescribed reverse diffusion process.

Flow matching offers an attractive alternative to diffusion models by learning a time-dependent velocity field that transports samples from a source distribution to the target distribution along a prescribed probability path~\cite{liu2023}. Its relatively straight transport trajectories can reduce the number of sampling steps required for image generation. Under a linear probability path, the velocity predicted at an intermediate point can be used to directly estimate the destination image. This property provides a flexible framework for incorporating data consistency into inverse problems~\cite{Kim_2025_ICCV,pourya2026flower, Martin2025PnPFlow, 11584907}. Recently, Flower~\cite{pourya2026flower} provides an approximate Bayesian framework that separates destination estimation, data-consistency refinement, and stochastic propagation. However, its data-consistency refinement assumes a Gaussian observation model and therefore does not directly account for the Poisson statistics of PET measurements.

In this study, we present flow matching-based PET image reconstruction methods. First, we establish PET-FlowDPS as a baseline by extending flow-driven posterior sampling (FlowDPS)~\cite{Kim_2025_ICCV} to PET reconstruction using Poisson likelihood guidance with preconditioned gradients. Second, inspired by the Flower framework~\cite{pourya2026flower}, we propose a model-based PET reconstruction method that uses a pretrained flow-matching model as a learned image prior. The proposed algorithm iteratively performs three steps: (1) flow-based destination estimation to obtain a clean image estimate, (2) penalized PET image reconstruction refinement to enforce data consistency, and (3) flow-based propagation to update the image to the next time step. These three steps can be interpreted within an approximate Bayesian framework. Separating data-consistency refinement from flow propagation allows established model-based PET reconstruction algorithms to be directly incorporated into the refinement step without modifying the pretrained flow model, thereby enabling stable data-consistency enforcement for quantitative PET reconstruction.

 \begin{figure*}[!t]
  \centering
  \includegraphics[width=\textwidth]{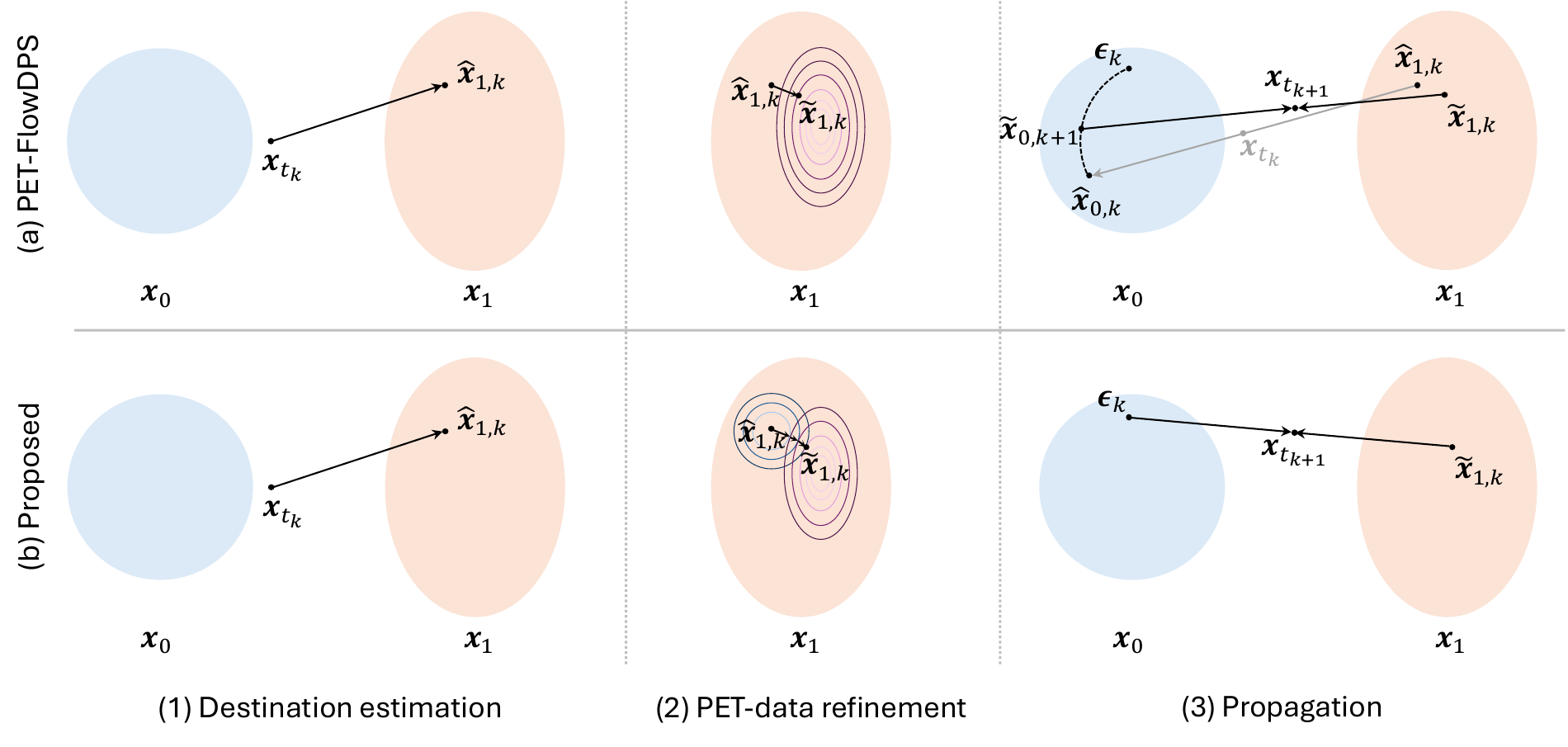}
  \caption{Overview of (a) PET-FlowDPS and (b) the proposed model-based PET reconstruction with a flow-based prior. In step (1), both methods estimate a destination image using the pretrained flow model. In step (2), (a) PET-FlowDPS applies a data-consistency update using the gradient of the log-likelihood, and (b) the proposed method performs a MAP reconstruction that balances data fidelity and the conditional flow-based prior. In step (3), the refined destination estimate is propagated to the next flow time point using the respective stochastic propagation rules. The purple contours in step (2) represent the PET log-likelihood, and the blue contours in (b-2) represent the conditional flow-based prior approximated by a Gaussian distribution.}
\label{fig:framework_overview}
\end{figure*}

\section{Background}
\subsection{PET image reconstruction}
PET image acquisition can be described using the following linear model
\begin{equation}
\bar{\bm{y}}=\bm{A}\bm{x}+\bm{b},
\label{eq:forward_model}
\end{equation}
where $\bar{\bm{y}}\in\mathbb{R}^{M}$ denotes the expected projection data, $\bm{x}\in\mathbb{R}^{N}$ represents the unknown radiotracer distribution, $\bm{A}\in\mathbb{R}^{M\times N}$ is the system matrix, and $\bm{b}\in\mathbb{R}^{M}$ represents mean of the random and scatter events. $M$ and $N$ denote the number of lines of response (LOR) and image voxels, respectively. Assuming the measured projection data $\bm{y}\in\mathbb{R}^{M}$ follows the Poisson distribution with mean $\bar{\bm{y}}$, the log-likelihood function can be written as
\begin{equation}
L(\bm{y}\mid\bm{x}) = \sum_{i=1}^{M} \left\{ y_i\log\left([\bm{A}\bm{x}]_i+b_i\right)
- \left([\bm{A}\bm{x}]_i+b_i\right) \right\}.
\label{eq:log_likelihood}
\end{equation}

The MLEM algorithm estimates $\bm{x}$ by maximizing \eqref{eq:log_likelihood} using the following iterative update for voxel $j$
\begin{equation}
x_{j}^{(n+1)} = \frac{x_j^{(n)}}{S_j} \sum_{i=1}^{M} A_{ij} \frac{y_i} {[\bm{A}\bm{x}^{(n)}]_i+b_i},
\quad
S_j=\sum_{i=1}^{M} A_{ij}.
\label{eq:mlem}
\end{equation}

This update rule is stable because it always satisfies $L(\bm y|\bm x^{(n+1)}) \geq L(\bm y|\bm x^{(n)})$.

\subsection{Flow matching}
Flow matching learns a continuous-time velocity field that transports samples from a simple source distribution to a target image distribution~\cite{Lipman2023,liu2023}. Let $\bm{x}_0\sim p_0(\bm{x}_0)$ denote a sample from the source distribution, typically a standard Gaussian distribution, and let $\bm{x}_1\sim p_1(\bm{x}_1)$ denote a sample from the target image distribution. Using a linear probability path, an intermediate sample at time $t\in[0,1]$ is defined as
\begin{equation}
\bm{x}_t = (1-t)\bm{x}_0+t\bm{x}_1,
\label{eq:linear_flow_path}
\end{equation}
and the corresponding conditional velocity along this path is
\begin{equation}
\bm{v}_t(\bm{x}_t\mid\bm{x}_0,\bm{x}_1) = \frac{d\bm{x}_t}{dt} = \bm{x}_1-\bm{x}_0.
\label{eq:conditional_velocity}
\end{equation}

A neural network $\bm{v}_{\bm{\theta}}(\bm{x}_t,t)$ parameterized by $\bm{\theta}$ is trained to approximate the velocity field by minimizing the conditional flow-matching loss
\begin{equation}
\mathcal{L}_{\mathrm{CFM}}(\bm{\theta}) = \mathbb{E}_{t,\bm{x}_0,\bm{x}_1} \left[ \left\| \bm{v}_{\bm{\theta}}(\bm{x}_t,t)
- (\bm{x}_1-\bm{x}_0) \right\|_2^2 \right],
\label{eq:flow_matching_loss}
\end{equation}
where $\bm{x}_t = (1-t)\bm{x}_0+t\bm{x}_1$, $t\sim\mathcal{U}(0,1)$,
$\bm{x}_0\sim p_0(\bm{x}_0)$, and $\bm{x}_1\sim p_1(\bm{x}_1)$. After training, samples from the target distribution can be generated by solving the ordinary differential equation (ODE)
\begin{equation}
\frac{d\bm{x}_t}{dt} = \bm{v}_{\bm{\theta}}(\bm{x}_t,t),
\qquad \bm{x}_0\sim p_0(\bm{x}_0),
\label{eq:flow_ode}
\end{equation}
from $t=0$ to $t=1$ using a numerical ODE solver.

Under the linear probability path in \eqref{eq:linear_flow_path}, an estimate of the destination image $\bm{x}_1$ can be obtained directly from an intermediate state $\bm{x}_t$ as
\begin{equation}
\hat{\bm{x}}_1(\bm{x}_t,t) = \bm{x}_t + (1-t)\bm{v}_{\bm{\theta}}(\bm{x}_t,t).
\label{eq:destination_estimate}
\end{equation}

This explicit destination-image estimate enables data-consistency refinement to be applied before advancing the intermediate state along the flow trajectory.

\section{Methodology}
A schematic overview of PET-FlowDPS and the proposed model-based PET reconstruction with a flow-based prior is shown in Fig.~\ref{fig:framework_overview}. Both methods refine the flow-based destination estimate using the measured PET data. PET-FlowDPS applies a data-consistency update using the gradient of the log-likelihood, whereas the proposed method performs a MAP reconstruction that balances data fidelity and the flow-based prior.

\subsection{FlowDPS for PET reconstruction}
FlowDPS is a posterior sampling solver for inverse problems that integrates the likelihood gradient and stochastic noise into a flow matching framework~\cite{Kim_2025_ICCV}. For the linear conditional flow, the marginal velocity is
\begin{equation}
\bm{v}_t(\bm{x}_t)=\frac{\bm{x}_t}{t}+ \frac{1-t}{t}\nabla_{\bm{x}_t}\log p_t(\bm{x}_t).
\label{eq:flow_velocity_score}
\end{equation}

The corresponding posterior marginal velocity is
\begin{equation}
\bm{v}_t(\bm{x}_t\mid\bm{y}) = \frac{\bm{x}_t}{t} + \frac{1-t}{t} \nabla_{\bm{x}_t} \log p_t(\bm{x}_t\mid\bm{y}).
\label{eq:posterior_velocity_score}
\end{equation}

Using Bayes' rule, \eqref{eq:posterior_velocity_score} can be rewritten as
\begin{equation}
\begin{aligned}
\bm{v}_t(\bm{x}_t\mid\bm{y}) &= \bm{v}_t(\bm{x}_t) + \frac{1-t}{t}\nabla_{\bm{x}_t}\log p_t(\bm{y}\mid\bm{x}_t) \\
&\approx \bm{v}_{\bm{\theta}}(\bm{x}_t,t) + \frac{1-t}{t} \nabla_{\bm{x}_t}\log p_t(\bm{y}\mid\bm{x}_t).
\end{aligned}
\label{eq:flowdps_posterior_velocity}
\end{equation}
Since $p_t(\bm y \mid \bm x_t)$ is not tractable, DPS~\cite{chung2023diffusion} and FlowDPS~\cite{Kim_2025_ICCV} apply the following approximations
\begin{equation}
\begin{aligned}
\nabla_{\bm{x}_t}\log p_t(\bm{y}\mid\bm{x}_t) 
&\approx \nabla_{\bm{x}_t} \log p\left( \bm{y}\mid\hat{\bm{x}}_1(\bm{x}_t,t) \right),
&& \text{(DPS)}
\\ &\approx \frac{1}{t} \nabla_{\hat{\bm{x}}_1} \log p\left( \bm{y}\mid\hat{\bm{x}}_1(\bm{x}_t,t) \right).
&& \text{(FlowDPS)}
\end{aligned}
\label{eq:flowdps_likelihood_approximation}
\end{equation}

Thus, the unconditional flow can be guided toward the posterior distribution by incorporating the likelihood gradient. Although DPS requires backpropagation through the flow model to compute the Jacobian of the destination estimator, FlowDPS approximates this gradient to bypass the Jacobian computation and applies the likelihood gradient directly to the destination image estimate.

To apply the likelihood approximation in~\eqref{eq:flowdps_likelihood_approximation}, FlowDPS uses the flow-based source and destination estimates. Using \eqref{eq:linear_flow_path} and \eqref{eq:conditional_velocity}, these estimates at each time point $t_k$ can be expressed by the flow-version of Tweedie's formula~\cite{Kim_2025_ICCV} as
\begin{align}
\hat{\bm{x}}_{0,k}
&= \bm{x}_{t_k} - t_k\bm{v}_{\bm{\theta}}(\bm{x}_{t_k},t_k),
\\ \hat{\bm{x}}_{1,k} &= \bm{x}_{t_k} + (1-t_k)\bm{v}_{\bm{\theta}}(\bm{x}_{t_k},t_k).
\end{align}
The destination estimate is then corrected using the likelihood gradient as
\begin{equation}
\tilde{\bm{x}}_{1,k} = \hat{\bm{x}}_{1,k} + \lambda_{t_k} \nabla_{\hat{\bm{x}}_{1,k}} \log p\left( \bm{y}\mid \hat{\bm{x}}_{1,k}  \right),
\label{eq:flowdps_destination_guidance}
\end{equation}
where $\lambda_{t_k}$ is the time-dependent guidance strength with the factor $1/t_k$ in \eqref{eq:flowdps_likelihood_approximation} absorbed into $\lambda_{t_k}$. After this correction, the sample at the next time point is obtained as
\begin{align}
\tilde{\bm{x}}_{0,k+1} &= \sqrt{t_{k+1}}\,\bm{\epsilon}_k + \sqrt{1-t_{k+1}}\,\hat{\bm{x}}_{0,k},
\\ \bm{x}_{t_{k+1}} &= (1-t_{k+1})\tilde{\bm{x}}_{0,k+1} + t_{k+1}\tilde{\bm{x}}_{1,k},
\end{align}
where $\bm{\epsilon}_k\sim\mathcal{N}(\bm{0},\bm{I})$.

For PET reconstruction, the preconditioned gradient of the log-likelihood is calculated by
\begin{equation}
\tilde{\nabla}_{\bm{x}} L(\bm{y}\mid\bm{x}) = \frac{\bm{x}}{\bm{S}} \left\{\bm{A}^{\mathsf{T}} \left[ \frac{\bm{y}} {\bm{A}\bm{x}+\bm{b}} \right] - \bm{S} \right\},
\label{eq:flowdps_pet_gradient}
\end{equation}
where ${\bm{x}}/{\bm{S}}$ is a preconditioning term for stability. Using this gradient, \eqref{eq:flowdps_destination_guidance} becomes
\begin{equation}
\tilde{\bm{x}}_{1,k} = \hat{\bm{x}}_{1,k} + \lambda \tilde{\nabla}_{\hat{\bm{x}}_{1,k}} L(\bm{y}\mid\hat{\bm{x}}_{1,k}),
\label{eq:pet_flowdps_destination_guidance}
\end{equation}
where $\lambda$ is the constant guidance strength. The overall algorithm of the PET-FlowDPS reconstruction is shown in Algorithm~\ref{alg:petflowdps}.

\begin{algorithm}[t]
  \caption{PET-FlowDPS reconstruction}
  \label{alg:petflowdps}
  \begin{algorithmic}[1]
    \REQUIRE Flow time points $0=t_0<t_1<\cdots<t_K=1$, flow model $\bm{v}_{\bm{\theta}}$, measured data $\bm{y}$, background $\bm{b}$, and guidance strength $\lambda$
    \STATE $\bm{x}_{t_0}\sim\mathcal{N}(\bm{0},\bm{I})$
    \FOR{$k=0$ {\bfseries to} $K-1$}
        \STATE $\hat{\bm{x}}_{0,k} = \bm{x}_{t_k} - t_k\bm{v}_{\bm{\theta}}(\bm{x}_{t_k},t_k)$
        \STATE $\hat{\bm{x}}_{1,k} = \bm{x}_{t_k} + (1-t_k)\bm{v}_{\bm{\theta}}(\bm{x}_{t_k},t_k)$
        \STATE $\begin{aligned} \tilde{\nabla}_{\hat{\bm{x}}_{1,k}} L(\bm{y}\mid\hat{\bm{x}}_{1,k}) =
        \frac{\hat{\bm{x}}_{1,k}}{\bm{S}} \left\{ \bm{A}^{\mathsf{T}} \left[ \frac{\bm{y}} {\bm{A}\hat{\bm{x}}_{1,k}+\bm{b}} \right] - \bm{S} \right\}
        \end{aligned}$
        \STATE $\tilde{\bm{x}}_{1,k} = \hat{\bm{x}}_{1,k} + \lambda \tilde{\nabla}_{\hat{\bm{x}}_{1,k}} L(\bm{y}\mid\hat{\bm{x}}_{1,k})$
        \STATE $\bm{\epsilon}_k \sim\mathcal{N}(\bm{0},\bm{I})$
        \STATE $\tilde{\bm{x}}_{0,k+1} = \sqrt{t_{k+1}}\,\bm{\epsilon}_k + \sqrt{1-t_{k+1}}\,\hat{\bm{x}}_{0,k}$
        \STATE $\bm{x}_{t_{k+1}} = (1-t_{k+1})\tilde{\bm{x}}_{0,k+1} + t_{k+1}\tilde{\bm{x}}_{1,k}$
    \ENDFOR
    \STATE \textbf{return} $\hat{\bm{x}}=\bm{x}_{t_K}$
  \end{algorithmic}
\end{algorithm}
Although FlowDPS derives an additional likelihood-gradient term for the posterior velocity field, its practical implementation differs from this theoretical formulation. In practice, the guidance coefficient is typically chosen to be a relatively small value, and multiple gradient updates are performed~\cite{Kim_2025_ICCV}. In addition, the stochastic propagation step is inherited from the sampling strategy of DPS~\cite{chung2023diffusion}, rather than being directly derived from the flow-based probabilistic formulation. As a consequence, the probabilistic interpretation of the trade-off between the pretrained flow-based prior and the data-consistency term becomes less explicit. The PET-FlowDPS algorithm improves the stability of the data-consistency update by incorporating an EM-based preconditioner. As a result, likelihood refinement can be performed stably using a single update. However, this modification still does not provide a probabilistic formulation of the overall reconstruction process.

\subsection{Model-based PET reconstruction using flow matching} 
In this section, inspired by the Flower framework, we propose a model-based PET reconstruction algorithm designed to improve stability and interpretability compared with PET-FlowDPS. The proposed reconstruction method formulates the flow-based prior, PET-data refinement, and stochastic propagation within an approximate Bayesian framework.

\begin{algorithm}[!t]
  \caption{Model-based PET reconstruction using flow matching}
  \label{alg:petflower}
  \begin{algorithmic}[1]
    \REQUIRE Flow time points $0=t_0<t_1<\cdots<t_K=1$, flow model $\bm{v}_{\bm{\theta}}$, measured data $\bm{y}$, background $\bm{b}$, penalty strength $\beta$, and inner iteration number $N$
    \STATE $\bm{x}_{t_0}\sim\mathcal{N}(\bm{0},\bm{I})$
    \FOR{$k=0$ {\bfseries to} $K-1$}
        \STATE $\hat{\bm{x}}_{1,k} = \bm{x}_{t_k} + (1-t_k)\bm{v}_{\bm{\theta}}(\bm{x}_{t_k},t_k)$
        \STATE $\bm{x}^{(0)} = [\hat{\bm{x}}_{1,k}]_{+}$
        \FOR{$n=0$ {\bfseries to} $N-1$}
            \STATE $\hat x^{(n+1)}_{\mathrm{EM},j} = \frac{x_j^{(n)}}{S_j} \sum_{i} A_{ij}\frac{y_i}{[\bm{A}\bm{x}^{(n)}]_i+b_i}$
            \STATE $\begin{aligned}
            x_j^{(n+1)}
            &=
            \frac{1}{2}
            \left(
            \hat{x}_{1,j,k}
            -
            \frac{S_j}{\beta}
            \right)
            \\
            &\quad+
            \frac{1}{2}
            \sqrt{
            \left(
            \hat{x}_{1,j,k}
            -
            \frac{S_j}{\beta}
            \right)^2
            +
            4\hat{x}_{\mathrm{EM},j}^{(n+1)}
            \frac{S_j}{\beta}
            }
            \end{aligned}$
        \ENDFOR
        \STATE $\tilde{\bm{x}}_{1,k} = \bm{x}^{(N)}$
        \STATE $\bm{\epsilon}_k \sim\mathcal{N}(\bm{0},\bm{I})$
        \STATE $\bm{x}_{t_{k+1}} = (1-t_{k+1})\bm{\epsilon}_k + t_{k+1}\tilde{\bm{x}}_{1,k}$
    \ENDFOR
    \STATE \textbf{return} $\hat{\bm{x}}=\bm{x}_{t_K}$
  \end{algorithmic}
\end{algorithm}
\subsubsection{Theoretical background}
We first describe the probabilistic modeling structure used to derive the proposed algorithm. The PET data observation process is defined by the following probabilistic forward model
\begin{align}
&\bm x_0 \sim p_0(\bm x_0) = \mathcal{N}(\bm 0, \bm I), \\
&\bm x_1 = \Phi_1(\bm x_0) \sim p_1(\bm x_1), \\
&\bm y \sim \mathcal{P}(\bm y | \bm A \bm x_1 + \bm b), 
\end{align}
where $\Phi_1$ is a transportation function from the source domain to the target domain and $\mathcal{P}$ stands for Poisson distribution.
The variables are generated sequentially as $\bm x_0 \rightarrow \bm x_1 \rightarrow \bm y$.

Our goal is to sample PET images from the posterior distribution $p(\bm x_1 | \bm y)$. We discretize the continuous time interval $t \in [0, 1]$ into $K+1$ time points and perform sequential sampling as follows:
\begin{align}
\bm x_{t_0} &\sim p(\bm x_{t_0}), \\
\bm x_{t_1} &\sim p(\bm x_{t_1}| \bm x_{t_0}, \bm y), \\
&\vdots \notag \\
\bm x_{t_K} &\sim p(\bm x_{t_K}| \bm x_{t_{K-1}}, \bm y),
\end{align}
where $0=t_0 < t_1 < \cdots < t_{K-1} < t_K=1$.
Since $\bm x_0$ can be sampled directly from $\mathcal{N}(\bm 0, \bm I)$, the remaining task is to sample
\begin{align}
\bm x_{t_{k+1}} \sim p(\bm x_{t_{k+1}} | \bm x_{t_k}, \bm y),
\label{eq:sequential_sampling}
\end{align}
for any $k \in \{0, 1, ..., K-1\}$. 
Direct sampling from this conditional distribution is difficult, so we introduce an approximate model for $p(\bm x_{t_{k+1}} | \bm x_{t_k}, \bm y)$.

\subsubsection{Approximate probabilistic models and sampling decomposition}
Following Flower~\cite{pourya2026flower}, we use the following approximations to decompose the conditional sampling process:
\begin{align}
\tilde{p}(\bm x_1 | \bm x_{t_k}) &= \mathcal{N}(\hat{\bm x}_1(\bm x_{t_k},t_k), \beta^{-1} \bm I), \label{eq:FM_prior}\\
\tilde{p}(\bm x_{t_{k+1}}| \bm x_1, \bm x_{t_k})
&= p(\bm x_{t_{k+1}}| \bm x_1) \notag \\ &= \mathcal{N}( t_{k+1} \bm x_1, (1-t_{k+1})^2 \bm I). \label{eq:stochastic_assumption}
\end{align}

The first approximation in \eqref{eq:FM_prior} assumes that, given $\bm x_{t_k}$, the destination image $\bm x_1$ follows a Gaussian distribution centered at the flow-based estimate $\hat{\bm x}_1(\bm x_{t_k},t_k)$. The parameter $\beta$ represents the precision of this distribution and controls the strength of the regularization imposed by flow matching.
The second approximation in \eqref{eq:stochastic_assumption} replaces the transition conditioned on both $\bm x_{t_k}$ and $\bm x_1$ with a stochastic transition conditioned only on $\bm x_1$. This formulation provides a probabilistic interpretation of the stochastic propagation step.

Next, we consider the conditional distribution $p(\bm x_{t_{k+1}} \mid \bm x_{t_k}, \bm y)$ in \eqref{eq:sequential_sampling}.
By marginalizing over $\bm x_1$, applying the conditional independence implied by the Markov property, {\em i.e.} $p(\bm x_{t_{k+1}} | \bm x_1, \bm x_{t_k}, \bm y)=p(\bm x_{t_{k+1}} | \bm x_1, \bm x_{t_k})$, and subsequently applying the approximations in \eqref{eq:FM_prior} and \eqref{eq:stochastic_assumption}, we obtain
\begin{align}
p(\bm x_{t_{k+1}} &| \bm x_{t_k}, \bm y) \notag \\
&= \int p(\bm x_{t_{k+1}} | \bm x_1, \bm x_{t_k}, \bm y) p(\bm x_1 | \bm x_{t_k}, \bm y) \mathrm d \bm x_1, \\
&\approx \int \tilde{p}(\bm x_{t_{k+1}} | \bm x_1, \bm x_{t_k}) \tilde{p}(\bm x_1 | \bm x_{t_k}, \bm y)\mathrm d \bm x_1,\label{eq:after_marginal_and_approximation}
\end{align}
where, by Bayes' rule and conditional independence implied by the Markov property, {\em i.e.} $p(\bm y | \bm x_1, \bm x_{t_k})= p(\bm y | \bm x_1)$,
\begin{align}
\tilde{p}(\bm x_1 &| \bm x_{t_k}, \bm y) \propto p(\bm y | \bm x_1, \bm x_{t_k}) \tilde{p}(\bm x_1 | \bm x_{t_k})= p(\bm y | \bm x_1) \tilde{p}(\bm x_1 | \bm x_{t_k}) \notag \\
&=\mathcal{P}(\bm y | \bm A \bm x_1 + \bm b)\mathcal{N}(\bm x_1|\hat{\bm x}_1(\bm x_{t_k},t_k), \beta^{-1} \bm I).
\label{eq:conditional_posterior}
\end{align}

Equation \eqref{eq:after_marginal_and_approximation} represents the approximate conditional distribution of $\bm x_{t_{k+1}}$ as a marginal distribution over $\bm x_1$.
Therefore, rather than sampling directly from this marginal distribution, we employ a hierarchical sampling. 
We first draw $\tilde{\bm x}_1$ from its conditional posterior and then draw $\bm x_{t_{k+1}}$ from the conditional distribution given $\tilde{\bm x}_1$. Specifically, it is given by
\begin{align}
&\tilde{\bm x}_1 \sim \tilde{p}(\bm x_1 | \bm x_{t_k}, \bm y), \label{eq:MAP_procedure} \\
&\bm x_{t_{k+1}} \sim \mathcal{N}(t_{k+1}\tilde{\bm x}_1, (1-t_{k+1})^2 \bm I). \label{eq:stochastic_procedure}
\end{align}

In Flower~\cite{pourya2026flower}, under a Gaussian likelihood, this sampling step was replaced by the expected value (or an equivalent MAP solution) plus a stochastic perturbation. Although a Laplace approximation could be used for approximate sampling from the Poisson posterior in PET, Flower reported better reconstruction performance when this perturbation was omitted. Based on this finding, our proposed method replaces step \eqref{eq:MAP_procedure} with the MAP solution:
\begin{align}
  \tilde{\bm x}_1 = \arg\max_{\bm x_1} \log \tilde{p}(\bm x_1 | \bm x_{t_k}, \bm y).
\end{align}

This MAP approximation converts the posterior refinement into a penalized likelihood optimization problem, which is well suited to PET reconstruction because efficient optimization algorithms, such as penalized EM methods, are readily available. Therefore, the proposed method uses the MAP estimate instead of directly sampling from $\tilde{p}(\bm x_1 \mid \bm x_{t_k}, \bm y)$ at each flow step.

\subsubsection{Update rules}
The proposed algorithm consists of three steps: (1) flow-based destination estimation to obtain a clean image estimate, (2) penalized PET image reconstruction refinement to enforce data consistency, and (3) flow-based propagation to update the image at the next time step.

\begin{enumerate}
\item Based on the conditional prior in \eqref{eq:FM_prior}, the destination image is estimated as the mean of the conditional
prior $\tilde{p}(\bm x_1 | \bm x_{t_k}) = \mathcal{N}(\hat{\bm x}_1(\bm x_{t_k},t_k), \beta^{-1} \bm I)$:
\begin{equation}
\bm{\hat{x}}_1(\bm{x}_{t_k}, t_k) = \bm{x}_{t_k} + (1-t_k)\,v_{\bm{\theta}}(\bm{x}_{t_k},t_k).
\label{eq:destination_estimation}
\end{equation}
\item Based on the conditional posterior in \eqref{eq:conditional_posterior}, the destination image estimate $\hat{\bm x}_{1,k}$ is refined using the measured PET data $\bm y$ by solving the following penalized PET reconstruction problem:
\begin{align}
\bm{\tilde{x}}_{1,k} &= \arg\max_{\bm{x}} \left\{ L(\bm{y}|\bm{x}) - \frac{\beta}{2} \left\|\bm{x}-\bm{\hat{x}}_1(\bm{x}_{t_k}, t_k)\right\|_2^2 \right\}, \notag \\
&= \text{prox}_{-\beta^{-1} L(\bm{y}|\bm{x}) }[ \bm{\hat{x}}_1(\bm{x}_{t_k}, t_k) ],
\label{eq:penalized_refinement} 
\end{align}
where $\beta$ controls the strength of the flow-based prior.
To solve \eqref{eq:penalized_refinement}, we use the optimization transfer method~\cite{Langed2000, Wang2012}. The voxel-wise update rule for (\ref{eq:penalized_refinement}) is
\begin{align}
\tilde{x}_{1,j}^{(n+1)} =& \frac{1}{2} \left[ \hat{x}_{1,j} - \frac{S_j}{\beta} \right] \notag\\
&+ \frac{1}{2} \sqrt{\left( \hat{x}_{1,j} - \frac{S_j}{\beta} \right)^2 + 4x_{\mathrm{EM},j}^{(n+1)}\frac{S_j}{\beta} },
\label{eq:closed_form_update}
\end{align}
where $\bm x_{\mathrm{EM}}^{(n+1)}$ is obtained by MLEM~\eqref{eq:mlem} from $\tilde{\bm x}_{1}^{(n)}$.
\item Based on the stochastic propagation model in \eqref{eq:stochastic_procedure}, the refined destination estimate is propagated to the next time point as
\begin{align}
\bm x_{t_{k+1}} = (1-t_{k+1}) \bm \epsilon_k + t_{k+1} \bm{\tilde{x}}_1,
\end{align}
where $\bm\epsilon_k \sim \mathcal{N}(\bm 0,\bm I)$ is sampled at each time step.
\end{enumerate}

The overall algorithm of the proposed model-based PET image reconstruction method is shown in Algorithm~\ref{alg:petflower}. To further improve reconstruction stability, we used an ensemble average of five independently reconstructed images, denoted as Proposed(5).

\begin{figure}[!t]
  \centering
  \includegraphics[width=0.94\linewidth]{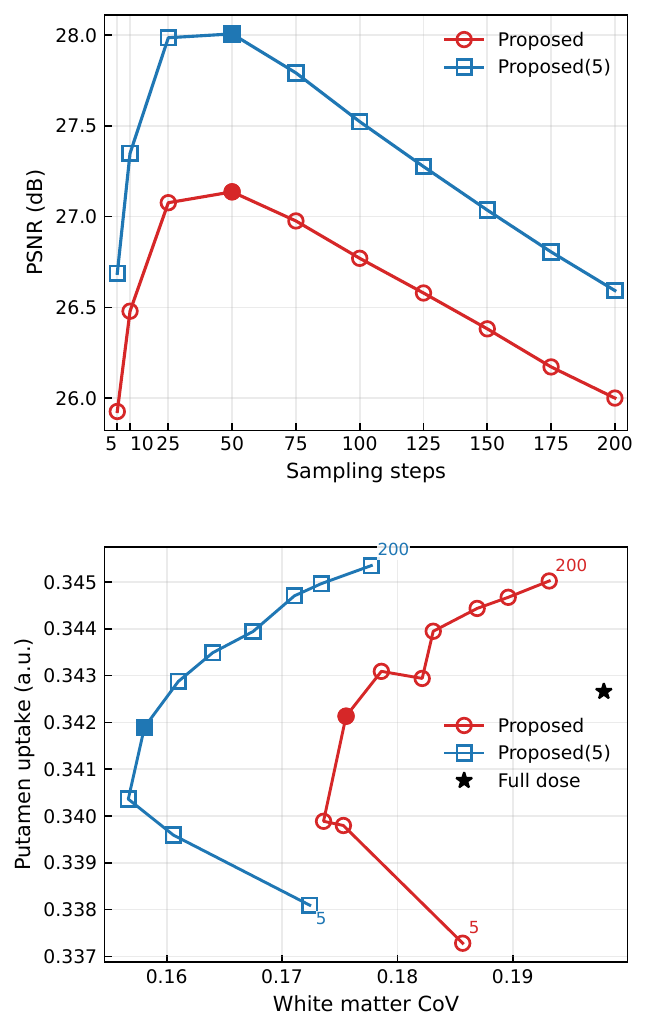}
  \caption{Impact of the number of sampling steps $K$ on PSNR (top) and on the putamen uptake--white matter CoV tradeoff curves (bottom). Filled markers correspond to the images as shown in Fig. \ref{fig:recon_results}.}
\label{fig:timesteps}
\end{figure}

\begin{figure}[!t]
  \centering
  \includegraphics[width=0.9\linewidth]{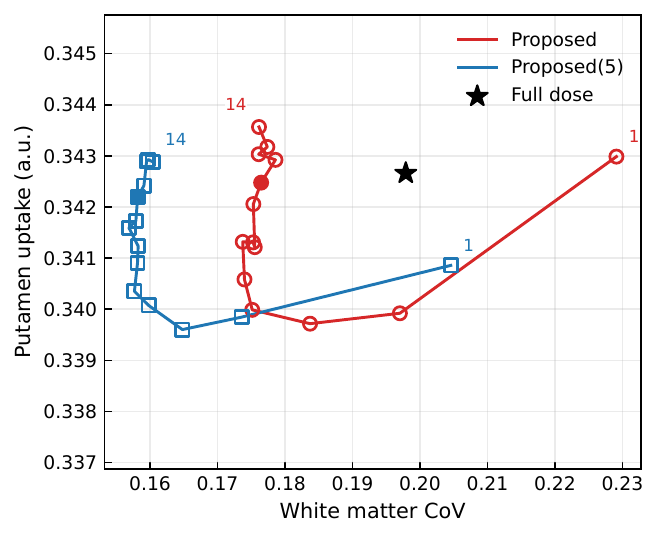}
  \caption{Impact of the number of inner EM iterations $N$ on the putamen uptake--white matter CoV trade-off curves across the 10 subjects at the 10\% count level. Markers were plotted for $N=1,2,\ldots,14$. Filled markers indicate $N=10$, which was used for the representative images shown in Fig.~\ref{fig:recon_results}.}
\label{fig:em_iterations}
\end{figure}

 \begin{figure*}[!t]
  \centering
  \includegraphics[width=\textwidth]{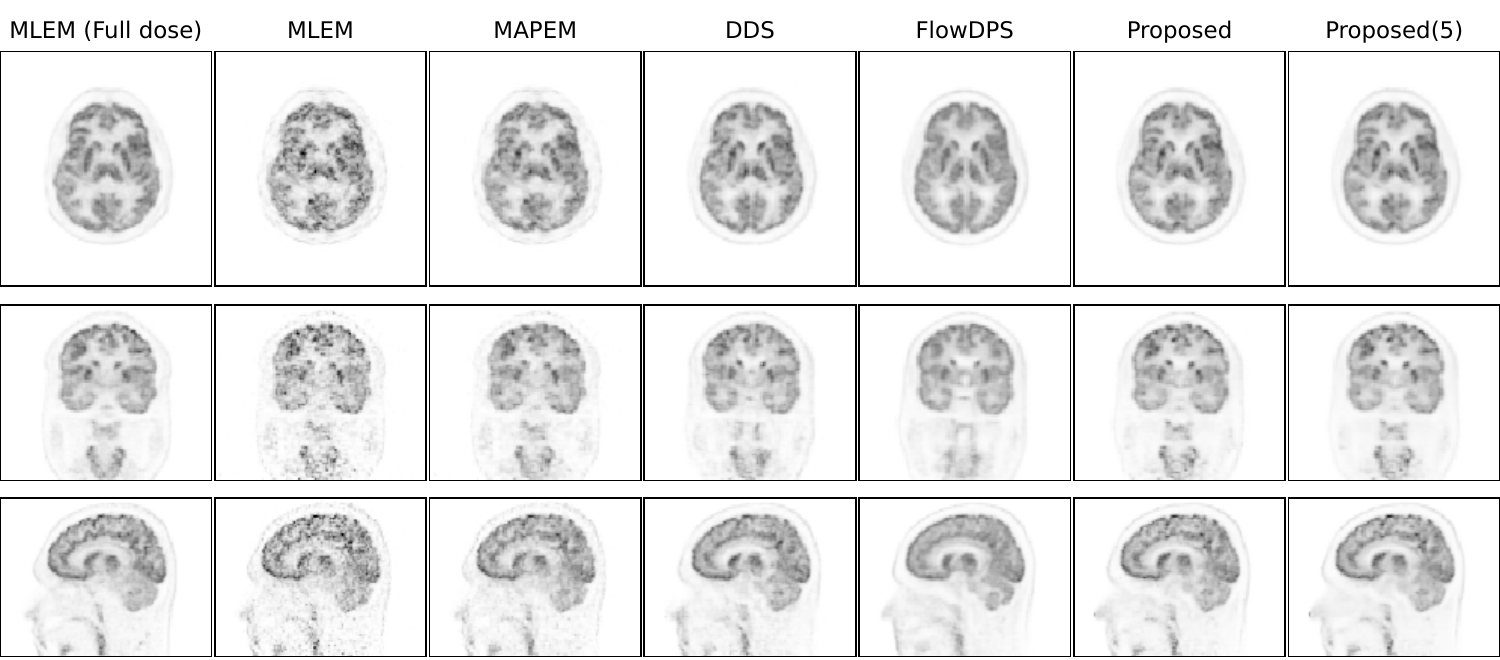}
  \caption{Reconstruction results of three orthogonal slices of the clinical [$^{\text{18}}\text{F}$]FDG data at 10\% dose.}
\label{fig:recon_results}
\end{figure*}

\begin{figure}[!t]
  \centering
  \includegraphics[width=\linewidth]{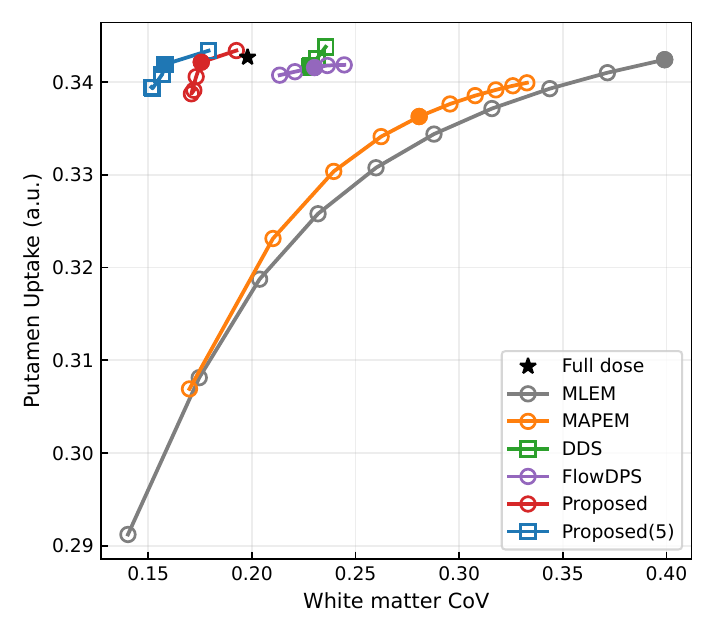}
  \caption{Mean putamen uptake--white matter CoV trade-off curves across
the 10 subjects at the 10\% count level. Markers were plotted at
10-iteration intervals from 10 to 100 for MLEM, at 20-iteration
intervals from 20 to 200 for MAPEM
($\beta_{\mathrm{MAPEM}}=30$, $\gamma=2$), at
$\lambda_\text{DDS}$ of 0, 10, 100, 500, and 1000 for DDS, at
$\lambda$ of 0.8, 0.9, 1.0, 1.2, and 1.4 for PET-FlowDPS, and at
$\beta$ of 0.05, 0.06, 0.08, 0.1, and 0.25 for Proposed and
Proposed(5). Filled markers correspond to the representative
images shown in Fig.~\ref{fig:recon_results}.}
\label{fig:tradeoff}
\end{figure}

 \begin{figure*}[!t]
  \centering
  \includegraphics[width=\textwidth]{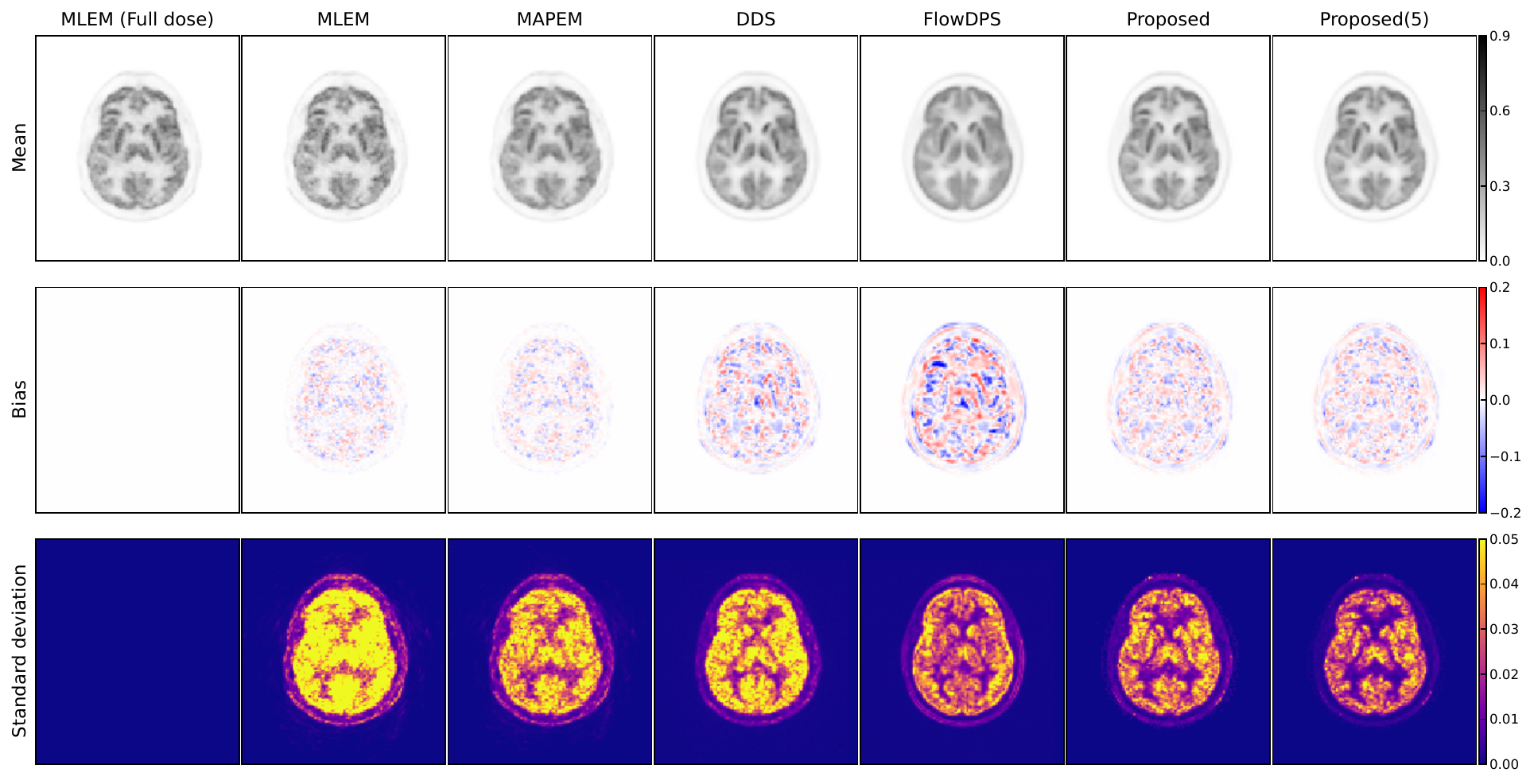}
  \caption{Voxel-wise mean, bias, and standard-deviation maps calculated from 10 independent 10\% low dose realizations for the different reconstruction methods.}
\label{fig:mean_bias_std}
\end{figure*}

\section{Experimental Setup}

We evaluated flow matching-based reconstruction methods using clinical brain $[^{18}\mathrm{F}]$FDG PET data. The experiments were run on an NVIDIA B200 GPU with 192 GB of memory.

\subsection{Brain PET data}
For the clinical data experiment, we used pretrained unconditional flow models trained on early 0-10 min frames of $[^{18}\mathrm{F}]$MK-6240 tau scans~\cite{Gong2024DDPM}, with an injected dose of approximately 185 MBq. These early-time-frame PET images mainly reflected cerebral blood flow information. The PET images were reconstructed using the ordered subsets EM algorithm with 3 iterations and 17 subsets, including point spread function modeling and time-of-flight information. The reconstructed images had a matrix size of $256 \times 256 \times 89$ and a voxel size of $ 1.17  \times 1.17 \times 2.79 $ mm$^3$. The PET images were subsequently downsampled and centrally cropped, resulting in a matrix size of $128 \times 128 \times 80$ and a voxel size of $2.34 \times 2.34 \times 2.79$ mm$^3$. The model was implemented using a 3D U-Net backbone and trained on 116 PET datasets. Among these datasets, 110 were used for training and 6 for validation.

During testing, we used 10 clinical $[^{18}\mathrm{F}]$FDG data from the Monash DaCRA fPET--fMRI dataset~\cite{Jamadar2022}. Dynamic PET scans were acquired over 90 min following an injection of approximately 238 MBq of $[^{18}\mathrm{F}]$FDG, and the data acquired during the 80--90 min interval were used for testing. Low dose PET data were simulated by downsampling the original list-mode data to 10\% and 5\% of the original counts. The 10\% dose level was used for the primary evaluation, and the 5\% dose level was evaluated to assess reconstruction performance under more severe low-dose conditions. The corresponding full dose data were used to generate reference images. The PET images were reconstructed on a $128\times128\times80$ voxel grid with a voxel size of $2.34\times2.34\times2.79$ mm$^3$. Scatter and random corrections were performed using a voxel-driven scatter model and a maximum-likelihood method, respectively, and magnetic resonance-derived $\mu$ maps were used for attenuation correction~\cite{Markiewicz2018}. The system matrix was implemented using Parallelproj~\cite{Schramm2024}.
\subsection{Evaluation}
We compared the proposed model-based PET reconstruction method with MLEM, MAPEM using the relative difference penalty~\cite{Nuyts2002}, decomposed diffusion sampling (DDS)~\cite{singh_2024}, and PET-FlowDPS. For a fair comparison, the diffusion model used in DDS was trained using the same training datasets and employed the same network architecture as the corresponding flow model.

For the MAPEM method, we used the relative difference penalty defined as
\begin{equation}
\label{eq:rdp}
R(\bm{x}) = \sum_{j}\sum_{k\in{N_j}}\frac{(x_j-x_k)^2}{(x_j+x_k)+\gamma\left|x_j-x_k\right|},
\end{equation}
where $N_j$ stands for a set of neighborhood indices of $j$th voxel and $\gamma$ controls the shape of the function. Following the default setting for clinical PET scanners, we set $\gamma=2$~\cite{Miwa2023}.

For DDS reconstruction, we used the following iterative update:
\begin{align}
\label{eq:dds}
\hat{\bm{x}}_0^{(n+1)}&=\hat{\bm{x}}_0^{(n)}\notag 
+\\
&\frac{\hat{\bm{x}}_0^{(n)}}{\bm{S}}\left[\bm{A}^T \frac{\bm{y}}{\bm{A}\hat{\bm{x}}_0^{(n)}+\bm{b}}-\bm{S}-2\lambda_\text{DDS}\left(\hat{\bm{x}}_0^{(n)}-\hat{\bm{x}}_0^{(0)}\right)\right],
\end{align}
where $\hat{\bm{x}}_0^{(0)}=\hat{\bm{x}}_0$, $n=0,1,\ldots,N_\text{DDS}-1$ denotes the number of iterations, $\hat{\bm{x}}_0^{(n)} / \bm{S}$ is a preconditioning term and $\lambda_\text{DDS}$ controls the strength of the regularization. We set the number of diffusion sampling steps to 200 and $N_\text{DDS}=5$ in the experiments.

For the proposed method, the flow interval was discretized into 50 sampling steps, and 10 penalized EM iterations were performed at each time point. The penalty strength was set to $\beta=0.1$. Unless otherwise specified, these settings were used for all reported results.

Mean uptake was calculated in the putamen region and background noise was evaluated using the coefficient of variation (CoV) in the white matter region as

\begin{equation}
\text{CoV} = \frac{\text{SD}_{\text{WM}}}{\text{Mean}_{\text{WM}}},
\label{eq:CoV}
\end{equation}
where $\text{Mean}_{\text{WM}}$ and $\text{SD}_{\text{WM}}$ are the mean and standard deviation values of the white matter regions, respectively. These regions were derived from FreeSurfer~\cite{FISCHL2012774}. Peak signal-to-noise ratio (PSNR) was also calculated using the full dose image as the reference as

\begin{equation}
\label{eq:psnr}
\text{PSNR} = 10\log_{10}\Biggl[\frac{\max{(\bm{X_\text{full}})}^2}{\frac{1}{N_\text{vox}}\bigl\lVert\bm{X_\text{full}}-\bm{X}'\bigr\rVert_{2}^{2}}\Biggr],
\end{equation}
where $\max{(\cdot)}$ indicates the maximum value of the image, $\bm{X_\text{full}}$ and $\bm{X}'$ denote the full dose and target reconstructed images, respectively, and $N_\text{vox}$ is the number of voxels.

For the variability analysis, 10 independent low-dose realizations were generated by randomly sampling 10\% of the events from the original listmode data. Each realization was independently reconstructed using all evaluated methods. Voxel-wise mean and standard-deviation images were calculated across the 10 reconstructed images. The bias image was calculated as the voxel-wise difference between the mean reconstructed image and the corresponding full-dose reference image.

\section{Results}
Figure~\ref{fig:timesteps} shows the impact of the number of sampling steps $K$ on PSNR and on the putamen uptake--white matter CoV tradeoffs. The PSNR peaked at $K=50$ and decreased thereafter. Favorable tradeoffs were observed with both 50 and 75 sampling steps. Figure~\ref{fig:em_iterations} presents the impact of the number of penalized EM iterations on the putamen uptake--white matter CoV tradeoff curves. Ten EM iterations ($N=10$) provided putamen uptake closest to the full dose reference while maintaining lower white matter CoV. Therefore, $K=50$ and $N=10$ were used in the subsequent experiments.

Figure~\ref{fig:recon_results} shows reconstructed images of one clinical $[^{18}\mathrm{F}]$FDG data using different methods. PET-FlowDPS and the proposed method produced images with lower noise levels than MLEM, MAPEM, and DDS. In particular, the proposed method provided visually higher image contrast than PET-FlowDPS, consistent with the putamen uptake--white matter CoV trade-off curves shown in Fig.~\ref{fig:tradeoff}. DDS, PET-FlowDPS and the proposed method recovered putamen uptake comparable to that of the full dose reference. However, DDS and PET-FlowDPS exhibited less favorable noise characteristics than the proposed method.

Figure~\ref{fig:mean_bias_std} shows the voxel-wise mean, bias, and standard-deviation maps for the different reconstruction methods.
MLEM, MAPEM, and DDS exhibited relatively high spatial standard deviations, whereas PET-FlowDPS and the proposed method showed lower spatial standard deviations. PET-FlowDPS exhibited relatively high spatial bias, whereas the proposed method maintained bias levels comparable to those of MLEM and MAPEM while providing lower spatial standard deviations. Overall, the proposed method provided a favorable balance between spatial bias and variability. 

\begin{figure*}[!t]
  \centering
  \includegraphics[width=\textwidth]{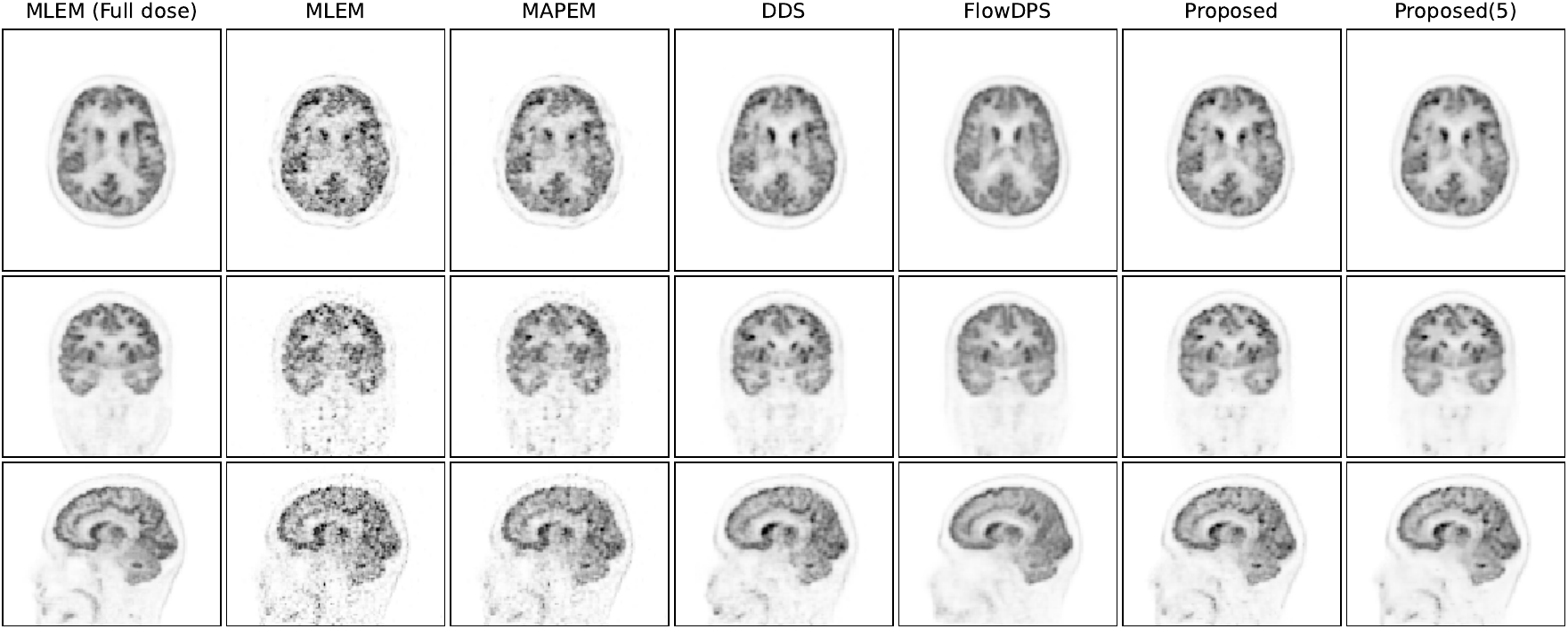}
  \caption{Reconstruction results of three orthogonal slices of the clinical [$^{\text{18}}\text{F}$]FDG data at 5\% dose.}
\label{fig:5precon_results}
\end{figure*}

\begin{figure}[!t]
  \centering
  \includegraphics[width=\linewidth]{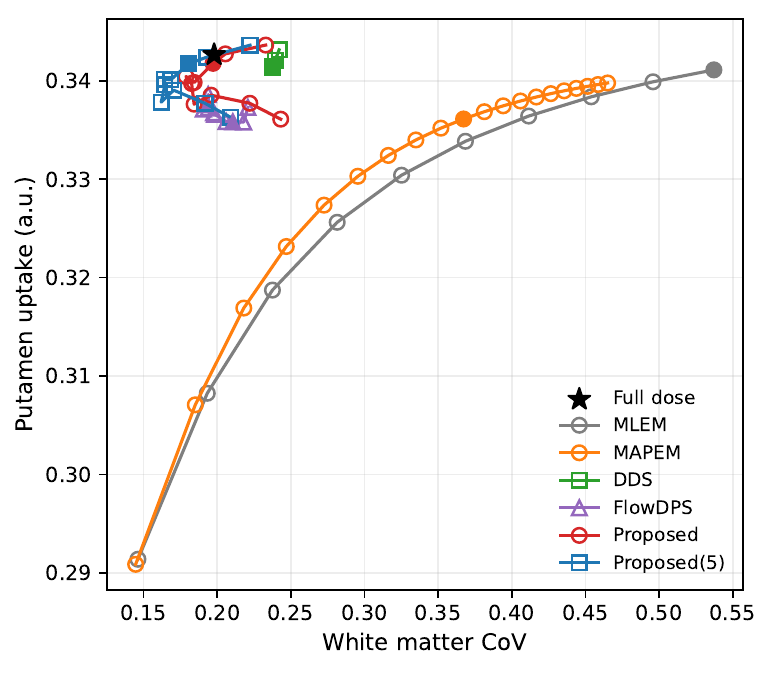}
  \caption{Mean putamen uptake--white matter CoV trade-off curves across the 10 subjects at the 5\% count level. Markers were plotted at 10-iteration intervals from 10 to 100 for MLEM, at 10-iteration intervals from 10 to 200 for MAPEM, at $\lambda_{\text{DDS}}$ of 0, 10, 100, 500, and 1000 for DDS, at $\lambda$ of 0.8, 0.9, 1.0, 1.2, 1.4, 1.6, 1.8, 2.0, 2.2, and 2.5 for PET-FlowDPS, and at $\beta$ of 0.005, 0.01, 0.02, 0.03, 0.04, 0.05, 0.06, 0.08, 0.1, and 0.25 for Proposed and Proposed(5). Filled markers correspond to the images as shown in Fig. \ref{fig:5precon_results}.}
\label{fig:tradeoff5p}
\end{figure}

To further evaluate reconstruction performance under more severe low-count conditions, Fig.~\ref{fig:5precon_results} shows the reconstructed images at the 5\% count level. Compared with the 10\% dose results, PET-FlowDPS produced blurred images, whereas the proposed method preserved image contrast and structural details. Consistent with the reconstructed images, the putamen uptake--white matter CoV tradeoff curves in Fig.~\ref{fig:tradeoff5p} showed that the proposed method recovered putamen uptake comparable to that of the full dose reference, whereas PET-FlowDPS underestimated putamen uptake. These quantitative results demonstrate that the proposed model-based flow matching reconstruction outperformed both conventional iterative and generative model-based reconstruction methods.

The computation times for DDS, PET-FlowDPS, and the proposed method were 3.93, 0.40, and 1.85 minutes, respectively.

\section{Discussion}
In this study, we presented flow matching-based PET image reconstruction methods. Although diffusion models have shown promising performance for PET reconstruction, incorporating data consistency into the reverse sampling process remains challenging and can lead to computationally intensive and numerically sensitive updates. In addition, the generative sampling and data-consistency updates are closely coupled through the prescribed reverse diffusion process. Flow matching provides a flexible framework for addressing these limitations because, under a linear probability path, the destination image can be directly estimated from an intermediate state. This property allows destination estimation, PET-data refinement, and flow propagation to be treated as separate steps. Based on this framework, we first developed PET-FlowDPS by incorporating PET likelihood guidance into FlowDPS. We then proposed a model-based reconstruction method in which the PET-data refinement step is formulated as MAP reconstruction.

A key difference between PET-FlowDPS and the proposed method lies in the PET-data refinement step (2) shown in Fig.~\ref{fig:framework_overview}. PET-FlowDPS directly applies a likelihood-based correction to the flow-based destination estimate. Although the PET-specific preconditioning term improves the stability of this update, its behavior still depends on the selected guidance strength $\lambda$ and the local log-likelihood gradient. In contrast, the proposed method replaces this gradient-based refinement with MAP reconstruction. This formulation provides an explicit balance between the flow-based prior and the PET log-likelihood through the penalty parameter $\beta$, making the relative contributions of the learned prior and measured PET data more directly interpretable. This modification may stabilize the PET-data refinement step in the proposed method, which may explain its higher image contrast, lower spatial bias, and lower noise levels compared with PET-FlowDPS.

We investigated the effects of the number of sampling steps and the number of EM iterations. The results suggest that increasing the number of sampling steps does not necessarily lead to improved reconstruction performance. As shown in Fig.~\ref{fig:em_iterations}, increasing the number of EM iterations was important for recovering quantitative uptake and reducing bias. This finding suggests that sufficient optimization of the subproblem in \eqref{eq:penalized_refinement} is important for quantitative PET reconstruction. With few EM iterations, the refined image may remain influenced by the flow-based prior and may not sufficiently reflect the likelihood term. However, the effects of these two parameters are not independent because the total number of EM iterations increases with the number of sampling steps. In the proposed method, $N$ EM iterations are performed at each of the $K$ sampling steps, resulting in a total of $K\times N$ EM iterations. Therefore, the decrease in PSNR observed beyond 50 sampling steps may not be solely due to the increased number of sampling steps, but may also reflect the effect of the increased total number of EM iterations. Future work will further investigate the individual and combined effects of these parameters.

At the 10\% count level, DDS, PET-FlowDPS, and the proposed method recovered putamen uptake comparable to that of the full-dose reference, indicating that generative priors can preserve quantitative information under low-dose conditions. As shown in Fig.~\ref{fig:mean_bias_std}, DDS exhibited higher spatial standard deviations than the flow-based methods. PET-FlowDPS provided lower spatial variability at the cost of relatively increased spatial bias. In contrast, the proposed method maintained bias levels comparable to those of MLEM, MAPEM, and DDS while achieving lower spatial standard deviations. These results show that the proposed method provides a more favorable bias--variability balance than the other methods, with both low bias and low variability contributing to improved quantitative accuracy. At the 5\% count level, PET-FlowDPS underestimated putamen uptake, whereas the proposed method maintained uptake comparable to that of the full-dose reference. This result suggests that the MAP-based PET-data refinement is more robust under noisier conditions than the direct likelihood-guidance update used in PET-FlowDPS.

This study has several limitations. First, the evaluation was performed using 10 clinical brain PET datasets at two low-count levels (10\% and 5\%). Further evaluation using larger cohorts, different scanners, and other radiotracers is needed to evaluate the generalizability of the proposed method. Second, the full dose reconstructed images were used as references. These reference images contain statistical noise and correction errors, which should be considered when interpreting these quantitative measures. Third, hyperparameters, such as the number of sampling steps, the number of inner EM iterations, and the penalty strength, may require adjustment for different conditions. Fourth, the formulation of the proposed method relies on several approximations for some probabilistic models. Therefore, the reconstructed images should not be interpreted as exact samples from the PET posterior distribution.

\section{Conclusion}
In this work, we proposed flow matching-based PET image reconstruction methods. Evaluations showed that the proposed flow-based method improved PET image quality compared with MLEM, MAPEM, DDS, and PET-FlowDPS methods. The results demonstrate that flow matching is a promising generative prior for PET image reconstruction, offering a favorable balance between quantitative accuracy and denoising performance.

\bibliographystyle{IEEEtran}
\bibliography{ref}

\end{document}